\documentclass[11pt,a4paper]{article}

\usepackage{jstyle}
\usepackage{tikz}
\usepackage{amsthm,mathrsfs,stmaryrd,bm}
\usepackage[vcentermath]{youngtab}
\usepackage{simplewick}
\usepackage{framed,pifont}

\usepackage{graphics} 
\usepackage{graphicx}
\usepackage{epic}
\usepackage{eepic}

\usepackage{feynmp}
\usepackage[retainorgcmds]{IEEEtrantools}
\usepackage{tensor}

\usepackage{latexsym}
\usepackage{amsmath}
\usepackage{wasysym}
\usepackage{amsfonts}
\usepackage{dsfont}
\usepackage{epsf}
\usepackage{epsfig}
\usepackage{amssymb,bm,epsf,epsfig}

\usepackage{color}
\definecolor{rougef}{rgb}{0.56,0,0}
\definecolor{vertf}{rgb}{0,0.5,0}
\definecolor{bleuf}{rgb}{0,0,0.8}
\definecolor{violetf}{rgb}{0.5,0,0.5}

\usepackage[vcentermath]{youngtab}

 \csname
@addtoreset\endcsname{equation}{section}

\def\3s{{s \choose 3}}
\def\4s{{s \choose 4}}
\def\5s{{s \choose 5}}
\def\6s{{s \choose 6}}

\def\12{\dfrac{1}{2}}
\def\fr{\frac}
\def\ft{\footnote}

\def\2{\ell_2}

\def\pr{\partial}

\def\scri{\mathscr{I}}

\def\be{\begin{equation}}
\def\ee{\end{equation}}
\def\bea{\begin{eqnarray}}
\def\eea{\end{eqnarray}}
\def\ba{\begin{array}}
\def\ea{\end{array}}
\def\bec{\begin{center}}
\def\ec{\end{center}}

\def\d{\delta} 

\def\e{\epsilon}

\def\m{\mu}
\def\n{\nu}

\def\r{\rho}

\def\vf{\varphi}
\def\O{\Omega}

\def\cP{{\cal G}}

\def\cL{{\cal L}}

\def\cP{{\cal P}}

\author[a, b]{Dario Francia}
\author[c]{and Carlo Heissenberg}

\affiliation[a]{Museo Storico della Fisica e Centro Studi e Ricerche E. Fermi, Piazza del Viminale 1, I-00184 Roma,
Italy}
\affiliation[b]{Roma Tre University and INFN Roma Tre, via della Vasca Navale, 84 I-00146 Roma, Italy} 
\affiliation[c]{Scuola Normale Superiore and INFN Pisa, Piazza dei Cavalieri 7, I-56126 Pisa, Italy} 
\emailAdd{dario.francia@roma3.infn.it, \\ \hskip 42pt carlo.heissenberg@sns.it}

\title{\centering
\huge{Two-form asymptotic symmetries and \\ scalar soft theorems}}

\abstract{We investigate the large gauge transformations of a two-form gauge field in four-dimensional Minkowski space. Our goal is to establish a connection between these asymptotic symmetries and the scalar soft theorems described by Campiglia, Coito and Mizera whereas the soft scalar mode should be interpreted in terms of its two-form dual counterpart.}


\begin{document}

\begin{fmffile}{diagram}

\maketitle


\section{Introduction and outlook} \label{sec: intro}
In Refs. \cite{Campiglia:2017dpg, Campiglia:2017xkp}, it was pointed out that the soft theorem for the emission of a scalar particle can be recast, to  leading order and at tree level, in the form of a  Ward identity for the corresponding $\mathcal S$-matrix
\be\label{Id_CCM}
Q^+ \mathcal S - \mathcal S Q^-=0\,,
\ee
with $Q^\pm$ suitable operators expressed in terms of creation and annihilation operators of external physical quanta. $Q^\pm$ can be split into their hard parts $Q^\pm_h$ and soft parts $Q^\pm_s$. 

In particular, in retarded Bondi coordinates in four-dimensional Minkowski space, the soft ``scalar charges'' can be expressed  in terms of the massless scalar mode $\varphi(u, r, z, \bar z)=b(u, z, \bar z)/r + o(\fr{1}{r})$ as follows:
\be
Q^+_s= \int_{S^2} b(u, z, \bar z) \Lambda(z, \bar z)\gamma_{z\bar z}dz d\bar z\,, 
\ee
where $\Lambda(z,\bar z)$ is an arbitrary function of the two angular coordinates $z$ and $\bar z$ on the unit sphere while $\gamma_{z\bar z}$ is the corresponding metric. The interpretation of this Ward identity in terms of an underlying symmetry, however, remained elusive. Indeed, differently from the analogous results holding for the case of soft particles with spin $s \geq 1$ \cite{Barnich_Revisited, Barnich_BMS/CFT, Barnich_Charge, Strominger_YM, Strominger_Invariance, Strominger_Weinberg, Strominger_QED, Campiglia:2014yka, Campiglia,  hsp_1, hsp_2, hsp_3}, (for a review see Ref.~\cite{Strominger:2017zoo}) for the case of soft scalars it is not clear \emph{a priori} what the underlying symmetry should be that is capable of explaining the conservation of the corresponding charges.  Similar considerations would apply to the soft theorems for pions considered in Ref.~\cite{Hamada:2017atr}.
	
 In this paper we propose a relation between the operators $Q^\pm_s$, in four space-time dimensions, and the Noether charges associated to the large gauge symmetries of a two-form gauge field, to be interpreted as propagating the same massless scalar degree of freedom (d.o.f.), in a dual picture.\ft{During the completion of this work the paper \cite{Campiglia:2018see} appeared, where it was indeed proposed to interpret the scalar soft theorems in terms of the asymptotic symmetries of its dual two-form field, much in the spirit of our present approach.} In this fashion Eq.~\eqref{Id_CCM} would appear to be naturally interpreted as the Ward identity arising from the large gauge symmetry of a two-form field. 
	
 We begin in Sec.~\ref{sec:2formasympt} by exploring the asymptotic symmetries and the corresponding charges for a two-form gauge field in the radial gauge, selected so as to keep appropriate falloff conditions. The latter are identified so as to be compatible with the equations of motion in that gauge and so as to provide a finite flux of energy per unit of retarded time at null infinity. In particular, we identify the on-shell-propagating radiative mode in the component $B_{z\bar z}=rC_{z\bar z}(u, z, \bar z)+ o(r) $. 

In Sec.~\ref{sec:duality} we recall the duality relation connecting scalar and two-form fields in $D=4$, and identify the relation between the propagating scalar mode $b (u,z,\bar z)$ and the two-form physical d.o.f. $C_{z\bar z}(u,z,\bar z)$ as 
$C_{z\bar z} = - \gamma_{z\bar z}b$. On the other hand, the asymptotic charge for the two-form residual symmetry takes the form
\be
\tilde Q^+ = -\frac{1}{r}\int_{S^2} C_{z \bar z} \gamma^{z\bar z}(\partial_z \e_{\bar z}- \partial_{\bar z}\e_z) dz d \bar z\,,
\ee
where $\e_z(z, \bar z)$, $\e_{\bar z}(z, \bar z)$ are arbitrary gauge parameters [subject to gauge-for-gauge transformations generated by a scalar function $\e(z,\bar z)$ on the unit sphere]. We thus propose to further identify
\be
Q^+_s = r \tilde Q^+\,,
\ee
namely
$
\Lambda\gamma_{z\bar z} = \partial_z \e_{\bar z}- \partial_{\bar z}\e_z
$.

The possibility to analyze the relation between asymptotic symmetries and soft theorems from the perspective of dual theories may be worth exploring in a number of additional contexts.
To begin with, one option to further check the duality that we propose in our work
would be to look for a soft theorem for the two-form, so as to see whether it involves
the same scalar charge.
Moreover, while already approached to some extent for the case of electromagnetic fields in $D=4$ \cite{Strominger-dual, Seraj-dual}, it would be interesting to reconsider from this vantage point the issue of higher-dimensional asymptotic symmetries for gravity and for higher spins. It is tantalizing to speculate that some symmetries may be better identified in a given dual description rather than in other, on-shell equivalent, pictures, a possibility that is conceivable on account of the typically nonlocal relation that connects two dual covariant descriptions of the same d.o.f. On the other hand, dualities are notoriously difficult to keep beyond the free level, which may still be sufficient to some extent when dealing with asymptotic states, but certainly provides a serious warning about the possible scope of conclusions that may be drawn from these type of analyses.

Coming back to soft scalars, let us also observe that, while our main focus in this work is on the four-dimensional case, the very existence of analogous duality relations between free massless scalars and $(D-2)$-forms in $D$ dimensions provides natural candidate explanations for the corresponding soft scalar charges identified in any even $D$ in Ref.~\cite{Campiglia:2017xkp}, while also possibly indicating the existence of analogous results in odd dimensionalities as well.

\section{Asymptotic symmetries for two-form gauge fields} \label{sec:2formasympt}

We consider the gauge field described by an antisymmetric rank-two tensor $B_{\m \n} \, = \, - B_{\n \m}$ subject to the reducible gauge transformation
\be
\d \, B_{\m \n} \, = \pr_{\, \m} \, \e_{\n} \, - \, \pr_{\, \n} \, \e_{\m}\, ,
\ee
where the linear dependences among the components of $\e_{\m}$ are encoded in the gauge-for-gauge symmetry  $\d\e_{\m} \, = \, \pr_{\, \m} \, \e$, where $\e$ is a scalar parameter. The gauge-invariant field strength is 
\be
H_{\, \m \n \r} \, = \, \pr_{\, \m} \, B_{\n \r} \, + \, \pr_{\, \r} \, B_{\m \n} \, + \, \pr_{\, \n} \, B_{\r \m} \, ,
\ee
while the Lagrangian and equations of motion are given by\ft{We are adopting the mostly-plus signature.}
\be
\cL \, = \,- \fr{1}{6} \, H_{\, \m \n \r} \, H^{\, \m \n \r}\, ,  \qquad \, \pr^{\, \m} \, H_{\, \m \n \r} \, = \, 0 \, ,
\ee
or equivalently, in components of $B_{\, \m \n}$,
\be \label{eomB}
\Box B_{\, \mu\nu} \, + \, \nabla_{\mu}\, \nabla^\rho \, B_{\, \nu\rho} \,  -\, \nabla_\nu \, \nabla^\rho \, B_{\, \mu\rho}\, =\, 0\,.
\ee

Our goal in this section is to investigate the asymptotic symmetries of this theory, much in the spirit of what can be done for the Maxwell theory and for (linearized) gravity (see {\it e.g.} Refs.~\cite{Strominger_YM, Strominger_QED, Strominger_Invariance, Campiglia, Strominger_Weinberg, Barnich_Revisited, Barnich_BMS/CFT, Barnich_Charge, Campiglia:2014yka}) or for higher spins \cite{hsp_1, hsp_2, hsp_3}\ft{Asymptotic symmetries for fields of mixed-symmetry are a much less explored subject. For the case of $p-$forms see \cite{Afshar:2018apx}.}. We adopt Bondi retarded coordinates $x^{\, \m} = (u, r, z, \bar{z})$ such that the Minkowski metric in $D=4$ is
\be \label{bondi-coord}
ds^2 = - du^2 - 2 du dr + r^2 \gamma_{z\bar{z}}\, dz d\bar{z}\,,
\ee
where $\gamma_{z\bar{z}}$ is the metric of the Euclidean two-sphere.

To begin with, we exploit the gauge-for-gauge symmetry to set $\e_r=0$, thus fixing the the scalar parameter $\e$, up to an $r$-independent but otherwise arbitrary function $\e (u, z, \bar{z})$. Then, we employ the gauge transformations 
\be
\delta B_{ru} = \partial_r \e_u \,,\qquad \delta B_{ri} = \partial_r \e_i\,,
\ee
to reach the ``radial gauge'' 
\be
B_{ru}=0=B_{ri}\, ,
\ee
where $x^i$, with $i=1,2$, stand for $z$, $\bar z$. This leaves a residual gauge freedom with parameters $\e_u(u,z,\bar z)$ and $\e_i(u,z,\bar z)$, and the gauge-for-gauge redundancy $\e  (u,z,\bar z)$. We may then further exploit the $u$ dependence of $\e (u,z,\bar z)$ to set $\e_u (u,z,\bar z)=0$. The result of this gauge-fixing strategy is the following: one is left with the gauge-field components
\be
B_{ui}(u,r,z,\bar z)\,,\qquad
B_{z \bar z}(u,r,z,\bar z)\,,
\ee
while still keeping the residual gauge parameters
\be\label{transform-gauge}
\e_i \, (z, \bar z)\,,
\ee
together with the residual gauge-for-gauge symmetry encoded in
\be
\e \, (z,\bar z)\,.
\ee
Expanding the equations \eqref{eomB} in the above gauge yields
\begin{align}
	& \partial_r D^j B_{ju} \, =\, 0\,,\\
	& \partial_r^2 B_{ui}+\frac{1}{r^2}\partial_r D^{j} B_{ij} \, =\, 0\,,\\
	& \partial_u \partial_r B_{ui} \, - \,  \frac{1}{r^2}\partial_u D^j B_{ij} \, - \, \partial_r^2 B_{ui}\, - \, \frac{\Delta-1}{r^2} B_{ui} \, - \, \frac{1}{r^2}D_iD^j B_{ju} \, =\, 0\,,\\ \label{r-eq}
	& 2\left(\partial_r - \frac{1}{r}\right)\partial_u B_{ij}  \, - \,  \frac{\Delta}{r^2}  B_{ij}
	\, + \, \left( \partial_r - \frac{2}{r} \right) \left(D_{[i}B_{j]u} \, - \, \partial_rB_{ij}\right) \, - \, \frac{1}{r^2}D_{[i}D^l B_{j]l} \, =\, 0\,,
\end{align}
where $D_i$ denotes the covariant derivative on the unit sphere and $\Delta = D^i D_i$ is the Laplacian on the unit sphere.

In order to impose consistent falloff conditions, we adopt two guiding criteria: we consider field configurations that radiate a finite energy per unit time across any spherical section $S_u$ of null infinity and we check compatibility with the free equations of motion to leading order as $r\to\infty$.

The finiteness of the energy flux at infinity imposes that the limit
\be\label{energyflux}
\cP\, (u) =
\lim_{r\to\infty}\int_{S_u}\gamma^{ij}\gamma^{jk} H_{uil}(H_{ujk}-H_{rjk}) r^{-2} d\Omega
\ee
be finite, hence indicating that both $B_{ij}$ and $B_{uj}$ should scale at most like $r$, as $r\to\infty$. Equation \eqref{r-eq} further suggests that $B_{ij}$ should scale precisely like $r$, thus saturating the energy bound, so that the leading component of $\partial_u B_{ij}$ is unconstrained on-shell.
Indeed, we find that the free equations of motion are solved to leading order as $r\to\infty$ by
\be\label{falloff-radiation}
B_{ui} = D^j C_{ij} \log r+ \cdots\,,\qquad
B_{ij} = r C_{ij}+ \cdots\,,
\ee
where $C_{ij}(u,z,\bar z)$ is an antisymmetric tensor on the sphere. In particular, this class of asymptotic solutions highlights $C_{z\bar z}$ as the single on-shell propagating d.o.f. carried by the two-form field being the only independent function of the leading solution space. Moreover,  it carries a finite amount of energy to null infinity encoded in 
\be
\cP\, (u) \, = \, \int_{S_u} \gamma^{ij}\gamma^{jk} \partial_u C_{il}\partial_uC_{jk}\, d \O,
\ee
as required.

The falloff conditions \eqref{falloff-radiation} are  invariant under any gauge transformation parametrized by \eqref{transform-gauge}, which we thus identify as providing the set of asymptotic symmetries of the theory.  We can compute the corresponding surface charge \cite{Barnich-Brandt, HS-charges-cov, Avery-Schwab}
\be\label{generic-charge}
\tilde Q^+ = \oint_{S_u} \kappa^{ur} r^2 d\Omega\,,
\ee
where the integration is performed on a sphere $S_u$ at fixed retarded time $u$ and for a large value of the radial coordinate $r$, while the Noether two-form \cite{Avery-Schwab} $\kappa^{\mu\nu}$ satisfies 
\be
\kappa^{ur} \, = \, \e_\mu H^{\mu ur} \, = \, \frac{1}{r^2}\e_i \gamma^{ij} H_{jru} \, = \, \frac{1}{r^2}\e_i \gamma^{ij}\partial_r B_{uj}\,.
\ee
Making use of the equations of motion we can further rewrite the charge as follows
\be\label{charge_two_form}
\begin{split}
	\tilde Q^+ & =  -\frac{1}{r}\oint_{S_u} \gamma^{ij}\gamma^{lk} D_{i}\e_l\, C_{kj}\,d\Omega \\
	& = -\frac{1}{r} \int \gamma^{z\bar z}(\partial_z\e_{\bar z}-\partial_{\bar z}\e_z)C_{z\bar z}\,dzd\bar z\, .
\end{split}
\ee
\newpage
\section{Duality and scalar charges} \label{sec:duality}

As is well known, a two-form gauge field  $B_{\mu\nu}$ in $D=4$ is dual, on shell, to a scalar field $\vf$ via  the relation $\ast dB = d\vf$, where $d$ is the exterior derivative and $\ast$ is the Hodge dual in $D=4$\ft{This duality is just the simplest realisation of a group-theoretical result that allows to identify irreps $T$ and $\tilde T$ of $SO(n)$ described by different Young diagrams, whenever the lengths of the first columns are in the relation $\ell_1 = n - \tilde \ell_1$, with all the other columns being equal. The duality between scalar and two form in $D=4$ corresponds to the identification between the singlet representation $\bullet$ and the antisymmetric rank-$2$ form $\tiny \young(\hfill,\hfill)$ of SO(2), {\it i.e.} the little group for massless particles in $D=4$.}; explicitly
\be
\frac{1}{2}\,r^2\, \gamma_{z\bar z}\, \epsilon_{\mu\nu\rho\alpha}\,\partial^\mu B^{\nu\rho} = \partial_\alpha \vf\,.
\ee
In components, we have
\be\begin{aligned}
	&\partial_r B_{z\bar z} = r^2\gamma_{z\bar z} \partial_r \varphi\,,\\
	&\partial_r B_{uz} = -\partial_z \varphi\,,\\
	&\partial_r B_{u\bar z} = \partial_{\bar z} \varphi\,,\\
	&\partial_u B_{z\bar z}+D_{[z}B_{\bar z]u} - \partial_r B_{z\bar z} = -r^2 \gamma_{z\bar z} \partial_u \varphi\,.
\end{aligned}\ee
Comparing with the falloffs for the two-form \eqref{falloff-radiation}, we see that these equations are compatible to leading order with the standard falloff condition for the massless scalar
\be
\varphi(u,r,z,\bar z) = \frac{b(u,z,\bar z)}{r}+\cdots\,,
\ee 
provided one identifies
\be\label{link}
b\, \gamma_{z\bar z} = - C_{z\bar z}\,.
\ee
This relation provides the desired connection between the on-shell d.o.f. $C_{ij}$ of the two-form field and the propagating component $b$ of the massless scalar. Equation \eqref{link} can be rewritten covariantly as $C_{ij} =i b\, \Omega_{ij}$, where $\Omega$ is the standard symplectic form on the Euclidean sphere.

Let us now compare ``charge'' operators arising from scalar soft theorems \cite{Campiglia:2017dpg} with the surface charges given by two-form asymptotic symmetries \eqref{charge_two_form}, in order to connect the former to the latter by means of the duality transformation. 
We recall that the soft part of the scalar charges can be expressed as
\be
Q^+_s= \int_{S^2} b(u, z, \bar z) \Lambda(z, \bar z)\gamma_{z\bar z}dz d\bar z\,, 
\ee
where $\Lambda(z,\bar z)$ is an arbitrary function of the two angular coordinates.
In view of Eq.~\eqref{link}, we propose to identify
\be\label{charge-link}
Q^+_s = r \tilde Q^+\,,
\ee
and correspondingly for the residual symmetry parameters
\be
\Lambda\gamma_{z\bar z} = \partial_z \epsilon_{\bar z} \, - \, \pr_{\bar{z}} \, \epsilon_{z}\,.
\ee
A puzzling, although not completely unfamiliar\ft{See {\it e.g.} \cite{Strominger-any-D, Conde-Mao1, Conde-Mao2}, where charges depending on some inverse power of the radial coordinate were considered and interpreted as connecting asymptotic symmetries to soft theorems.} feature of the identification is the fact that, while the action of $Q^+_s$ is well defined on $\mathscr I^+$, {i.e.} even after performing the limit $r\to\infty$, our two-form asymptotic symmetry charge appears to vanish in the large-$r$ limit. This seems to be a consequence of the fact that, in radial gauge, symmetry parameters are not allowed to grow with $r$, and hence are unable to compensate for the falloff $\partial_r B_{ui}\sim 1/r$. 

A possible way out of this inconvenience could be to add terms of the type 
\be\label{Coulombs}
\partial_zf(z)\,r \,,\qquad
\partial_{\bar z} g(\bar z)\,r \, ,
\ee 
to the two-form components $B_{uz}$, $B_{u\bar z}$ respectively. These terms are indeed allowed by the leading equations of motion and give no contribution to the energy flux at infinity. These new terms would give rise to the modified charge
\be \label{modified charge}
\int \big[\epsilon_z \partial_{\bar z}g(\bar z)+\epsilon_{\bar z} \partial_zf(z)\big]
dzd\bar z
-\frac{1}{r} \int \gamma^{z\bar z}(\partial_z\epsilon_{\bar z}-\partial_{\bar z}\epsilon_z)C_{z\bar z}\,dzd\bar z\,,
\ee
which no longer goes to zero as $r\to\infty$. On the other hand, in this limit, it appears to become independent of the physical d.o.f. $C_{z\bar z}$ and, in the dual interpretation, of the radiative mode $b$ of the massless scalar. Indeed, the $f(z)$ and $g(\bar z)$, appearing in the first term of Eq.~\eqref{modified charge}, are related by duality to a scalar field $\varphi(u,r,z, \bar z) = -f(z)+g(\bar z)+\cdots$, which is static to leading order as $r\to\infty$.

The terms \eqref{Coulombs} admit a natural interpretation if one phrases the problem of studying the two-form falloffs in a spacetime of generic dimension $D$. In this extended setup, the asymptotic analysis of the equations of motion highlights two classes or ``branches'' of solutions: denoting by $x^i$ coordinates on the celestial $(D-2)$-sphere, one has a radiation branch
\be\label{radiation-branch}
B_{ui} =  \frac{2}{4-D} U_{i}(u,x^k)\, r^{(4-D)/2}+\cdots\,,\qquad
B_{ij} = C_{ij}(u,x^k)\, r^{(6-D)/2}+\cdots\,,
\ee	
subject to $U_i=D^j C_{ij}$ (unless $D=6$, in which case only $\partial_u U_i = \partial_u D^j C_{ij}$ need be imposed)
and a Coulomb-like branch
\be\label{Coulomb-branch}
B_{ui} = \tilde U_{i}(x^k) r^{5-D} +\cdots\,,\qquad
B_{ij} = \frac{1}{D-4}\tilde C_{ij}(u,x^k)r^{5-D}+\cdots\,,
\ee
where $\partial_u \tilde C_{ij} = D_{[i}\tilde U_{j]}$ and $(D-5)D^j \tilde U_{j}=0$. Solutions of the first type give rise to nonzero energy flux across sections of null infinity, $\mathcal P(u)\neq 0$, and only give vanishing contributions to the (global) charges as $r\to\infty$. The second class, on the other hand, does not contribute to the energy flux, while giving nonzero contributions to charge integrals. In $D=4$,  the above expressions exhibit singularities and Eq.~\eqref{radiation-branch} reduces to Eq.~\eqref{falloff-radiation}, while Eq.~\eqref{Coulomb-branch} gives rise to Eq.~\eqref{Coulombs}. 

The scalar radiative mode $b$, in four dimensions, appears thus to be dual to the a radiation solution for its two-form counterpart. From this observation, it appears natural that its soft charge may be dual to an asymptotically vanishing two-form charge. 

\acknowledgments

We would like to thank G.~Barnich and A. Campoleoni for discussions and comments and Ronak M. Soni for exchanges. We are grateful to the D\'epartement de Physique,  Physique Math\'ematique des Interactions Fondamentales Universit\'e Libre de Bruxelles for the kind hospitality extended to us during the completion of this work. The work of D.F. is supported by the Grant ``HIGHSPINS---Higher Spins and Their Symmetries'' of Museo Storico della Fisica e Centro
Studi e Ricerche E. Fermi. The work of C.H. was supported in part by Scuola Normale Superiore. The work of both of us was also supported in part by Istituto Nazionale di Fisica Nucleare-Iniziativa Specifica-GSS-Pi.\\

\emph{Note added.}---Athough the general logic is similar, our results are not identical to those of Ref.~\cite{Campiglia:2018see}. In particular, as already noticed, our charge vanishes in the limit $r \to \infty$ and in this sense it fails to reproduce the expected transformations on $\scri$. A possible source of explanation for this behavior may be looked for in the difference between the radial gauge, that we employ in this work, and the Lorenz gauge that is used in Ref.~\cite{Campiglia:2018see}, with the former possibly being amenable to being weakened by allowing for subleading corrections in the large-$r$ behavior of the radial components of the field $B_{\, \m \n}$. 
With hindsight, this may be taken as an indication that gauge independence of
the results is to be assumed with some care, when it comes to comparing asymptotic analyses,
in the absence of a general criterion allowing to establish \emph{a priori} which components of
the gauge transformations may contribute to a nonvanishing charge at $\mathscr I$, given
a set of consistent falloffs.




\end{fmffile}
\end{document}